# A Comprehensive Survey on Computer-Aided Diagnostic Systems in Diabetic Retinopathy Screening

Meysam Tavakoli[a*], Patrick Kelley[b]

[a]*Radiation Oncology Dept., University of Texas Southwestern Medical Center, Dallas, TX, USA*

[b]*Dept. of Physics, Indiana University-Purdue University, Indianapolis, IN, USA*

*Correspondence's email address:* meysamtavakkoli@gmail.com  (Meysam Tavakoli)

**Summary**

Diabetes Mellitus (DM) can lead to significant microvasculature disruptions that eventually causes diabetic retinopathy (DR), or complications in the eye due to diabetes. If left unchecked, this disease can increase over time and eventually cause complete vision loss. The general method to detect such optical developments is through examining the vessels, optic nerve head, microaneurysms, haemorrhage, exudates, etc. from retinal images. Ultimately this is limited by the number of experienced ophthalmologists and the vastly growing number of DM cases. To enable earlier and efficient DR diagnosis, the field of ophthalmology requires robust computer aided diagnosis (CAD) systems. Our review is intended for anyone, from student to established researcher, who wants to understand what can be accomplished with CAD systems and their algorithms to modeling and where the field of retinal image processing in computer vision and pattern recognition is headed. For someone just getting started, we place a special emphasis on the logic, strengths and shortcomings of different databases and algorithms frameworks with a focus on very recent approaches.

**Important Acronyms**

- **Diabetes Mellitus**: DM.

- **Diabetic Retinopathy**: DR.

- **Non-proliferation Diabetic Retinopathy**: NPDR. • **Proliferation Diabetic Retinopathy**: PDR.

- **Microaneurysm**: MA.

- **Haemorrhage**: HE.

- **Exudate**: EXs.

- **Cotton Wool Spots**: CWS.

- **Neovascularization**: NV.

- **Macular edema**: ME.

- **Field-of-View**: FOV.

- **Computer-Aided Diagnosis**: CAD.

- **Optic Nerve Head**: ONH.

- **Receiver Operating Characteristic**: ROC.

- **Area Under the Curve**: AUC.

- **k-Nearest Neighbor**: kNN.

- **Support Vector Machine**: SVM.

- **Neural Network**: NN.

- **Artificial Intelligence**: AI.

- **Machine Learning**: ML.

- **Deep Learning**: DL.

- **Convolution Neural Network**: CNN.

- **Recurrent Neural Network**: RNN.

- **Autoencoder**: AE.

# Contents

# 1. Introduction

The rapidly rising Diabetes Mellitus (DM) is one of the major issues of our recent health care [1]. The number of people affected by DM continues to increase at an alarming rate [2, 3]. The World Health Organization estimates for the next 25 years the number of diabetic people to grow from 130 to 350 million [4]. In the same direction, in the time between 2000 to 2030 another source anticipates the spread of the DM population worldwide to increase from 2.8% to 4.4% [5]. The real catch is that just half of the diabetic people are noticed about the disease [6]. From medical aspect viewpoint, DM leads to serious late complications such as macro and micro vascular changes which cause Diabetic Retinopathy (DR), heart disease, and renal problems [7, 8]. In DM, because of insulin insufficiency there is incapacitated metabolism of glucose; this in turn leads to damages and complications in blood vessels [9]. According to a study in the United States, DM is among the five lethal diseases, and yet there is no treatment for it. In the same study, the total yearly expense to diagnose and treat DM in 2002 was estimated to be $132 billion [10]. In fact, unmanaged DM and its complication leads to different malfunctions and death.

DR is a serious issue stemming from DM that affects the eye and is a main cause of vision loss and blindness [11, 12]. DR is the resulting condition specifically for these complications and damaged vessels lying in the tissue behind the retina [13]. Unfortunately so many cases of diabetes goes undiagnosed and the connection between diabetes and eye vision isn't common knowledge; patients with DM are 25 times more likely at risk of vision loss when compared with non-diabetic ones [14]. DR is often overlooked and detected too late, since it is a silent disease, and it may be diagnosed by the patient when the vascular abnormalities in the retina have progressed to a stage that its cure is complicated and sometimes unfeasible [15, 16]. Therefore, the most successful therapy for DR can be managed just in the early phases of that. Consequently, early detection of DR thru systematic screening is of critically important.

It is believed that the screening of diabetic people for the evaluation of DR potentially reduces the risk of blindness in these people by 50% [17]. The problem is however that, this would produce immense costs, due to the large number of examinations. Moreover, there are not enough specialists especially in rural areas to do these examinations [18]. Physical screening of DR to address the large and mounting number of DR cases is therefore not possible. Instead, automated systems based on image processing methods may help to overcome this issue [19]. It would be more beneficial and helpful if the initial steps of analyzing the retinal images and detecting all retinal lesions in the images can be computerized through a robust computer automated system. In this way, the automated system would allow for speedier identification of the telltale signs of DR from the early onset of retinal lesions and notify the ophthalmologist for further assessment. This would allow more patients to be screened per year and for the ophthalmologists to better manage and direct their time on those priority patients who would have otherwise been overlooked or delayed in treating [20].

*1.1. Clinical Signs of Diabetic Retinopathy*

Now, before we start DR diagnosis the big question is *what are the revealing indications of DR?* There are several abnormalities in the retina that are manifestations of DR which are concisely described below.

1) Microaneurysms (MAs): They are the first visible sign of DR. They appear as small and round shape dots near tiny retinal blood vessels in retinal images [21]. The size of MAs usually ranges from $10 \mu m$ to $125 \mu m$ in diameter [22, 19].

2) Haemorrhages (HEs): A more severe level of DR is caused by retinal haemorrhages or leakage of weak capillaries and ruptured MAs into the surface of the retina [23]. They are defined as a red spot with different shapes, such as "dot", "blot" and "flame" [23, 21] and irregular margin and uneven density [24].

3) Hard Exudates (EXs): When the lipoproteins and other proteins are leaking because of the degradation or breakdown of the blood-retina barrier through abnormal retinal vessels, hard EXs appear [25]. They become visible as small yellowish-white patches with shrill boarders [24]. They are often collocated in bulks or semicircular hoops and placed in the exterior layer of the retina [24].

4) Cotton Wool Spots (CWS): Also called soft exudates, they occur due to occlusion, or blockage of blood supply, of arterioles and cause disruption and damage to areas of tissue [26]. The ischemia of the retinal layer of nerve fiber caused by reduced blood flow to the retina changes the axoplasmic flow and leads to accumulation of axoplasmic residue in the retinal ganglion cell axons. The accumulation of this axoplasmic residue arise as fleecy yellowish-white lesions in the retinal nerve fiber layer known as Soft Exudates or CWS [26, 27].

5) Neovascularizations (NVs): The new blood vessels start to grow on the inner surface of the retina because of primarily the lack of oxygen. These new blood vessels push into the surrounding retinal space and are feeble and very often bleed into retinal cavity, effecting the eyesight [10, 28].

6) Macular edema (ME): It is recognized as the swelling of the macula, an area at the center of the retina. It is resulted in penetrance of abnormal capillaries in the retinal because of the fluid leakage or solutes around the macula [29, 30]. Eventually, it influences the central the eyesight [31].

Figures 1 and 2 show typical normal retinal images with its basic landmarks, and abnormal images with different signs and stages of DR.

*1.2. Stages of Diabetic Retinopathy*

According to the existence of the aforementioned of clinical signs, and the severeness/density of these signs [34] DR is categorized into four kinds including: mild, moderate, severe Non-Proliferative DR (NPDR), and PDR [35, 36, 24]. Respect to number of MAs and HEs, the NPDR is introduced as normal, mild, moderate, and sever DR [37].

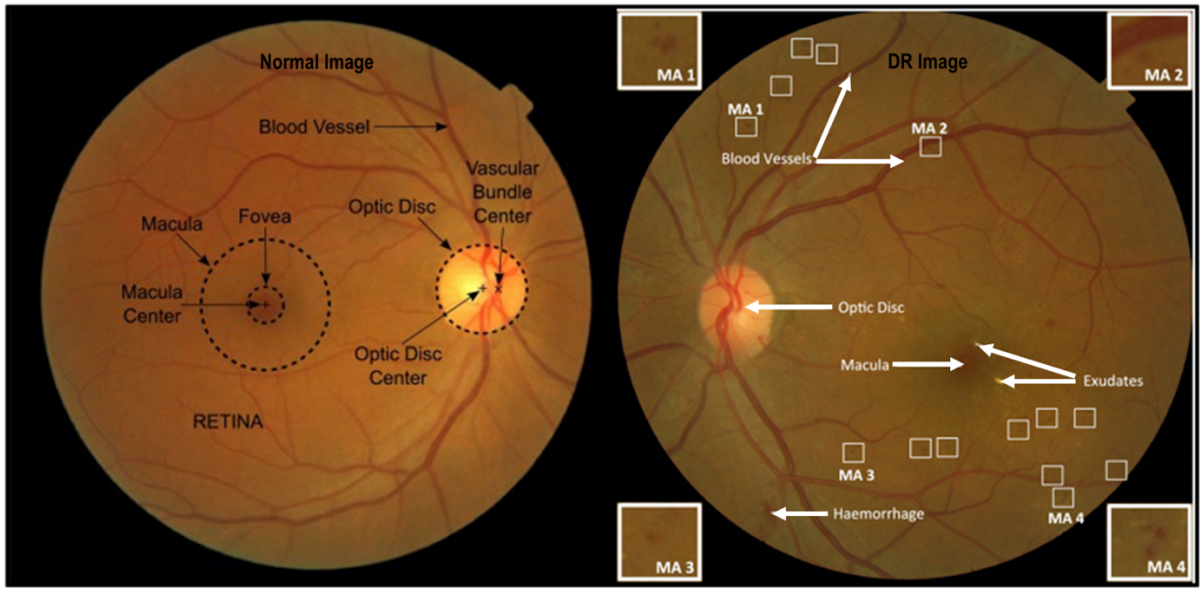

Figure 1: **Normal retinal image vs. DR one.** Sample retinal images for both normal and abnormal ones. On the left image, all available anatomical landmarks have been shown including fovea, macula, blood vessels, optic nerve head [32]. On the right one: there are all pathologic signs of DR including MAs, HEs, and EXs are visible [33].

- **Normal**: If no DR sign is observed.

- **Mild**: Here, only MAs are presented.

- **Moderate**: In this case, both MAs, and HEs are exist and their numbers are less than 20 in each quadrant, also we see soft EXs as well.

- **Severe**: Here, in each quadrant there are more than 20 MAs, and HEs. Also, in more than two quadrants there are some venous beadings and different shapes of hard EXs.

Figure 2 shows the retinal images of these categories and related signs. In the PDR, we see the initiation of new vessels, which are abnormal vessels (NVs). It is the severe phase of DR, in which these new abnormal vessel growth, caused by poor circulation especially from formations of MAs, protrude into the retina and even into the vitreous, the gel-like medium that provides the eye its spherical shape. Thus, both MAs and NVs are two important clinical lesions [38] and fluid in DR is classified as EXs and non-EXs [39].

There are several organizations for DR grading such as American academy of ophthalmology, which classification of DR was established by curing of DR at its early stage [40], and protocol introduced y Scottish DR grading system [41]. Table 1 presents Scottish protocol.

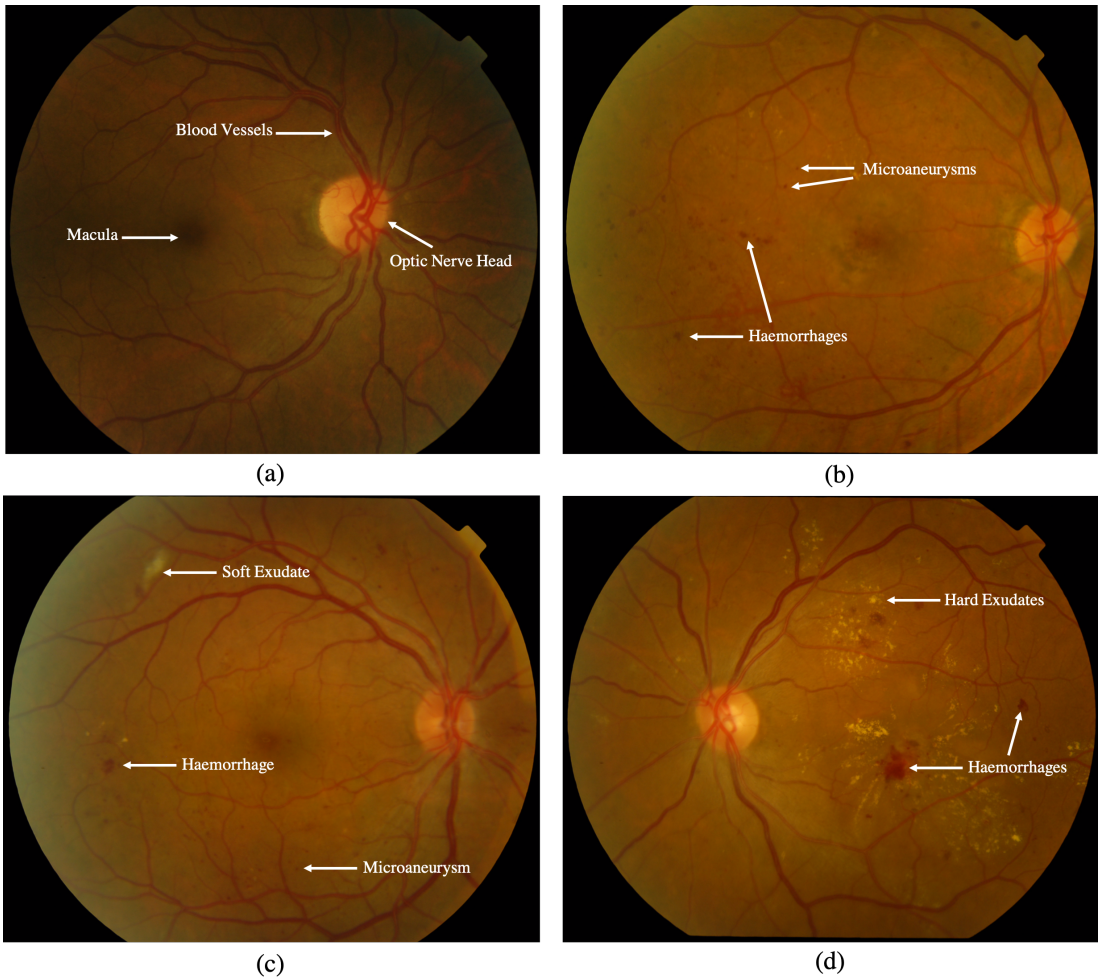

Figure 2: **Different types of DR.** (a) Normal; (b) Mild NPDR; (c) Moderate NPDR; (d) Severe NPDR.

Table 1: DR Scottish grading protocol [41].

| Grade | Characteristics |
|---|---|
| No DR | Without abnormalities |
| Mild NPDR | Just MAs appear |
| Moderate NPDR | There are more than 5 MAs but less than severe NPDR |
| Severe NPDR | - In each quadrant, there are more than 20 HEs |
| | - In two quadrants, there is Venous beading |
| | -Intra-retinal microvessel abnormalities |

| PDR | -Any NVs at ONH or other parts of the retina |
| :---: | :---: |
| | -Vitreous/ pre-retinal HE |
| No ME | There are not any EXs or thickening of retina in posterior pole |
| Mild ME | There are EXs or thickening of retina at posterior pole, >1 disc diameters from fovea |
| Moderate ME | Same as mild ME's signs plus 1 disc diameters, without affecting fovea |
| Severe ME | There are EXs or thickening of retina with affecting fovea |

MAs, in general, are very good indication of possibly worsening conditions of DR. From statistics point of view, the NPDR type with MAs has 6.2% chance to expand into PDR during a year [1]. An increased number of MAs is a main primary characteristic of DR progression. Moreover, with development of ischemia, there is an increased chance of PDR progress during first year. The first year risk growth increases from 11.3% (lower) to 54.8% (advanced) phase [1, 42]. In the NVs stage, diabetics have 25.6% to 36.9% chance of vision loss and blindness, if not cured appropriately. Furthermore, PDR eyes not therapied after more than 2 years have a chance of vision loss and blindness at 7.0% and if for more than 4 years it is not cured, the chance of vision loss goes up to 20.9%. On the other hand, this vision loss and blindness decreases to 3.2% during 2 years and 7.4% during 4 years of therapy [42]. Diabetic people with mild DR do not need specific treatment. In fact, they need to manage their DM and the related risk factors such as hypertension, anaemia, and renal malfunction. They require to closely be monitored, to decrease possibility of progressing higher stages of DR [18]. In the severe and advanced phase of DR, like NV, therapy is limited [43].

*1.3. Retinal Imaging and Databases*

Now that we are aware of the signs and stages of DR, the way diagnosis and determination of the stage of DR involves processing retinal images. Retinal photography is a procedure where 3-dimenssional retinal landmarks is projected to the 2-dimenssion image. The created image shows the amount of reflected light [44]. In the camera which create the fundus image, the optical layout is equivalent to ophthalmoscope, which is used to observe the inner part of the eye. The camera sees the retinal area with 2.5 times magnification at the angle between $30^{\circ}$ to $50^{\circ}$. More information can be found in [1]. In retinal photography for DR screening, the below modalities/techniques are utilizes [44].

1) Red-free retinal imaging: The image is taken by applying the amount of reflected light at a particular waveband.

2) Color retinal image: This image is based on Red Green Blue (RGB) spectrum and the camera detector sensitivity to the light.

3) Fluorescein angiography retinal image: The created image is based on the amount of photons that is emitted by fluorescein dye. The dye is injected through the blood stream [44].

Automated algorithms, in the process of testing and improving upon themselves, requires a set of exemplary data with the proper classifications. In our case, these would be retinal images from reputable databases, many of which have been classified by professional ophthalmologists. Most of the provided retinal images in these databases are used in application of screening and testing the proposed methods related to detection of MAs, HEs, exudates and other DR features. Outlined below, many of these online research databases are included. Table 2 summarizes all databases for detection of DR.

- **DIARTDB**: The dataset was created by Kauppi et al. [45] and is freely available for research from http://www.it.lut.fi/project/imageret/. Images were taken with a 50 degree field-of-view (FOV). It has two classes, DIARTDB0, including 130 color retinal images (20 images are normal without sign of DR and 110 contain symptoms of the DR). These signs include hard and soft EXs, MAs, HEs and NV. The second class is called DIARTDB1, including 89 images which 5 of them are normal and 84 ones contain at least mild NPDR signs (MAs).

- **HRF**: It is presented as the **H**igh-**R**esolution **F**undus retinal image dataset. The images were captured with different image acquisition setting and 45 degree FOV. In this database for each category of healthy, glaucomatous, and DR there are images. The information in database was created by a group of experts from the ophthalmology clinics [46].

- **STARE**: it consists of 400 images were initially applied by Hoover and Goldbaum [47]. These images were captured at 35 degree FOV.

- **KAGGLE**: The images were prepared by the Kaggle [37] as a series of high resolution ones. The images were captured under the different spatial conditions and were indexed by a well-experienced ophthalmologist respect to without DR to Mild, Moderate, Severe and Proliferative DR.

- **ROC**: **R**etinopathy **O**nline **C**hallenge is a dataset established for providing research groups to establish CAD system for diagnosis DR. It aimed to create a platform for algorithms evaluation [48].

- **DRIVE**: It has been developed from a DR screening system. The database included 40 retinal images in which 33 images without symptoms of DR and 7 ones showed signs of DR. The database was taken using a non-mydriatic 3CCD fundus camera with a 45 degree FOV [49].

- **MESSIDOR**: It consisted of 1200 eye retinal color images. The images were captured by a color non-mydriatic retinography with a 45 degree FOV with different resolutions [50].

- **E-OPTHIA**: It was created because of a DR screening project which developed a color dataset for DR, called E-ophtha. This dataset was made of fundus images with variety of signs such as EXs and MAs which manually labelled by well-experienced ophthalmologists [51]. It contained 47 images with EXs, 35 with no sign, and 281 images with MAs or small HEs.

- **DRIONS**: An evaluating dataset for ONH segmentation that included 110 retinal images. The fundus images were captured with a color analogical camera [37].

- **CHASEDB1**: It is a dataset available at blogs.kingston.ac.uk/retinal/chasedb1 for free. It included 28 fundus images. An ophthalmologist and human experts were asked to mark the vessels in all these images and these hand-labelled vessels were used as the ground truth (or gold standard) for this dataset [52].

- **HEI-MED**: The database is a test database for evaluating algorithms in the EXs and macular edema detection. It contained 169 images which were segmented manually by a high skilled ophthalmologist [30].

- Others: These are other datasets established by individual researchers groups such as MUMS-BD [19, 53] DMED, ARIA, Eyepack [1]

*Table2: More details for the databases use for DR detection system*

| Database | Numbers | Resolution | Format | Task |
|---|---|---|---|---|
| MESSIDOR [50] | 1200 | 1440x960, 2240x1488, 2304x1536 | TIFF | -DR grading -Risk of DME |
| Kaggle [54] | 80000 | - | JPEG | -No sign of DR -Mild -Moderate -Severe -PDR |
| E-ophtha[55] | 148 MAs, 233 normal, 47 EXs, 35 no signs of DR | 2544x1696 1440x960 | JPEG | -MA small HE detection -EX detection |
| DRIVE [49] | 33 normal 7 mild DR stage | 584x565 | TIFF | -Vessels extraction |
| STARE [47] | 402 | 605x700 | PPM | -13 retinal diseases |
| | | | | -Vessels segmentation -ONH |
| DIARETDB1[45] | 5 no sign<br>84 NPDR symptom | 1500x1152 | PNG | -MAs -HEs |
| CHASE [37, 56] | 28 | 1280x960 | JPEG | -Vessels extraction |
| MUMS [19, 53] | 120 | | PNG | MAs |
| ARIA [57] | 16 normal<br>92 AMD- 59 DR | 768x576 | TIFF | -ONH -Fovea -Vessel |
| SEED-DB [58] | 192 normal<br>43 glaucomatous | 3504x2336 | - | ONH |
| EyePACS [59] | 9963 | - | - | -Referable DR<br>-MA |
| ORIGA [60] | 482 normal<br>168 glaucomatous | 720x576 | - | ONH |

Table 3 summarizes some CAD systems at different image database analysis for DR detection. These approaches are distinguished based on features, processing, and classification of DR.

*Table 3: Some CAD systems and databases(Blue=B, Green=G, Red=R)*

| Author | Image plan | Method | database |
|---|---|---|---|
| Pourreza et al. [53] | B & G | Radon transformation | DRIVE, STARE, MUMS-DB |
| Esmaeili [61] | R& G | Morphology | STARE, DRIVE, DIARETDB1 |
| Qureshi et al. [62] | G | Entropy and Hough | DIARETDB0, DIARETDB1, DRIVE |
| Welfer et al. [63] | LUV | Mathematical Morphology | DRIVE, DIARETDB1 |
| Harangi and Hajdu [64] | Grey | Probabilistic model and augmented Naive based | DIARETDB0, DIARETDB1, DRIVE, MESSIDOR |
| Akyol et al. [65] | G | BLP | DIARETDB1, DRIVE,ROC |
| Dai et al. [66] | Grey | Hough, PCA | MESSIDOR, DRIONS |
| Alshayeji et al. [68] | Grey | Adaptive edge detection, | DERIVE, DMED, STARE, DIARETDB1 |
| Fraz et al. [68] | G | Watershed transformation | DRIVE, Shifa, CHASEDB1, DIARETDB1 |

## 2. Computer-Aided Diagnosis and Diabetic Retinopathy

As previously mentioned, manual detection and diagnosis is an exhaustive task in both cost and lack of experts [16] when screening DR. It would be more cost effective and helpful if the initial task of analyzing the retinal images can be automated. Automated approaches address these issue by decreasing the time, expense, and attempt considerably [69]. Moreover, since image processing techniques are growing in all areas of medical science, by assisting them, especially in advanced ophthalmology, we can do automated screening [39].

For this reason, automated DR detection and classification utilizes Computer Aided Diagnostic (CAD) systems. These computer-based systems have the ability to detect any change in normal and abnormal images and systematize these variations to form a feature space. At the end, the mixture of these features introduces type and stage of DR. Different CAD systems have been presented in state of arts for early DR detection and related lesions [12, 18, 19, 70-77].

Normally, each CAD system is the sequence of two different approaches i.e. feature detection and a classification [37]. Since there are many different approaches in CAD systems, it is also necessary to evaluate their robustness and accuracy. There are many different standard methods used in assessing the effectiveness of the CAD systems. A common one is using Receiver Operating Characteristic (ROC) analysis and concept of area under the curve (AUC) analysis. The AUC of ROC curve is a measurement of performance for automated method, for example, extraction problem, at different thresholds values. ROC is a probability curve and AUC shows degree or rate of distinguishability. In fact, AUC shows us how much the method is able to distinguish between classes. Higher the AUC, better the method is in its prediction. By analogy, higher the AUC, better the method is at distinguishing between images with DR and no DR. ROC curves illustrate the tradeoff between sensitivity and specificity for a range of thresholds and enable the identification of an optimal value [78]. Hence, assessment using the ROC curve is a way to evaluate the model, independent of the choice of a threshold.

Here, the analytical definition when using ROC is assessed according to the true positive fraction (TPF), given by sensitivity, and the false positive fraction (FPF), (1-specificity) [16]. Moreover, the accuracy is specified as a quantification including the ratio of well-classified pixels. The final outputs for the automated approach as compared to the ground truth or gold standard are calculated for each image. These metrics are defined as:

$$Sensitivity(Se) = \frac{TP}{TP+FN}$$
$$Specificity(Sp) = \frac{TN}{TN+FP}$$
$$Accuracy(Acc) = \frac{TP+TN}{TP+FN+TN+FP} \qquad (1)$$

where TP is true positive, TN is true negative, FP is false positive and FN is false negative same as in [79, 80].

A look into the results for both ROC and AUC (Eq. 1) obtained by others CAD systems in DR screening. Table 4 summarizes some of the algorithms at different image analysis for detection of DR.

*Table 4: Some of the CAD systems with different image analysis stages*

| CAD system | Methodology | Authors |
|---|---|---|
| ONH detection and segmentation | | Qureshi et al. [62], Welfer et al. [63], Harangi and Hajdu [64], Akyol et al. [65], Dai et al. [66], Alshayeji [67], Basit and Fraz [68], Bharkad [81], Lesay et al. [82], Zou et al. [83], Xiong and Li [84], Sarathi et al. [85], Lu [86], Lowel et. [87], Hoower et. [88] |
| Vessel segmentation | Thresholding approach, | Hoover et al. [47], Dash and Bhoi [89], Zhu et al. [90], Neto et al.[91], Xu and Luo [92], Nguyen et al. [93], Zhang et al. [94], Fraz et al. [95], Marin et al. [96], Akram et al. [97], Vermeer et al. [98], chakraborti et al. [99], Wu et al. [100], Tavakoli et al. [101], Roychowdhury et al.[102], Tolias and Panas [103], Azzopardi et al. [104], Zana et al. [105, 106], soares et al. [107] |
| | Tracking approach, | |
| | Machine classifiers | |
| Lesion detection | Red lesions (MAs, HEs) | Quellec et al. [108], Walter et al. [22], Tavakoli et al. [19], Wang et al. [109], Srivastava et al. [110], seoud et al [23], Pereira et al. [113], Dai et al. [111], van Grinsven et al. [112] |
| | Hard and soft exudates | Liu et al. [114], Zhang et al. [116], Orlando et al. [115], Youssef and Solouma [117], Fraz et al. [118], Amin et al. [119], sopharak et al. [120], Sanchez et al. [121] |
| DR detection and classification | DR, Non DR NPDR, PDR | Tavakoli et al. [19], Koh et al. [122], Gardner et al.[123], Rubini and Kunthavai [124], Gulshan et al. [125], Kose et al. [126], Yang et al. [127], Sisodia et |

| | | al. [128] , Kumar et al. [129], Gupta et al.[130], Leontidis [82], Sinthanayothin et al. [131], Usher et al. [132], Fleming et al. [17, 133], Perumalsamy et al. [134], Winder et al. [135], Mookiah et al. [136] |
|---|---|---|

Briefly, both Tavakoli et al. [19] and Philip et al. [36] introduced a DR screening system based on the comparison of automated system respect to the gold standard which was manual detection by ophthalmologist. Their approach detected DR images with an accuracy of more than 90%. Sinthanayothin et al. [131] established a trustable automatic DR online screening system purposed to simplify ophthalmologists' work in their country. They got sensitivity and specificity for their screening program to classify the DR 80.2% and 70.7% respectively. Usher et al. [132] employed DR screening tool which detected DR with both sensitivity and specificity of 95.1% and 46.3% respectively. Niemeijer et al. [70] introduced a DR detection algorithm based on information fusion approach with AUC of 0.88. Quellec et al. [71] established an automated DR screening system for both MAs and age-related macular degeneration detection, and their method reached to an AUC of 0.93. Fleming et al. [133] introduced a CAD system to detect blot HEs and provided with both sensitivity and specificity of 98.60% of 95.50% respectively. Fleming et al. [17] also established another CAD system to detect MAs which got a sensitivity and specificity of 85.4% and 83.1%. Perumalsamy et al. [134] employed a CAD system and obtained an accuracy of 81.3% by evaluating the performance with ophthalmologists' idea. Abramoff et al. [137] established a web-based DR detection system with specific protocol and short questionnaire, measurement of visual eyesight, and four retinal images. The accuracy of their protocol reached 93%. Abramoff et al. [138] also presented an automated DR detection system which acquired an AUC of 0.84 of DR detection.

There are also some authors that reviewed most of the CAD methods, and their applications [1, 39, 44, 135, 139] for automated screening of DR. Winder et al. [135] discussed different segmentation approaches of Optic nerve head (ONH), retinal vessels, macula, and DR detection. Teng et al. [139] for retinal landmarks, and lesions segmentation, and image registration reviewed different methods. Abramoff et al. [44] surveyed various retinal imaging modalities, vessel segmentation, lesion, and ONH detection. Patton et al. [39] discussed image preprocessing, and registration methods, besides landmarks and lesions segmentation. Mookiah et al. [1] comprehensively reviewed the approaches to detect and extract the retinal lesions and landmarks such as ONH, vessels, EXs, MAs, and HEs.

From classification viewpoint, color characteristics were utilized on Bayesian classifier to group each pixel into either lesion or non-lesion categories [140]. In this study, authors obtained 100% accuracy in detection of all EXs in the retinal images, and accuracy of 70% in classification of normal retinal images. By using image processing techniques and multilayer Neural Network (NN), DR and normal retina were categorized with sensitivity and a specificity of 80.2% and 70.7% respectively [21]. Automated detection of NPDR, based on MAs, HEs, and EXs was studied in [136, 141]. The approach was able to correctly detect the NPDR stage with an accuracy of 81.7%. For DR screening again above lesions were used [142] and obtained the sensitivity and specificity of 74.8% and 82.7%, respectively. An early automated DR detection system was developed by Kahai et al. based on decision support system [143]. In their method, Bayes optimality criteria was used to identify DR with a sensitivity and specificity of 100% and 67%. Different stages of DR were grouped using both area and geometry of the combination of the vessels with a feedforward NN [144]. The average classification efficiency for this method was 84% with both sensitivity and specificity of 90% and 100%. Nayak et al. employed EXs and vessel area along with parameters related to its texture which coupled with NN to classify NPDR, PDR, and normal [5]. They got accuracy of 93%, sensitivity 99%, and specificity of 100%. The feature of support vector machine (SVM) classifier also used by Acharya et al. [145] to classify the retinal image in to normal, mild, moderate, severe and PDR categories. They showed a classification accuracy of 82% with sensitivity of 82%, and specificity of 88%. Nicolai et al. established an automated lesion detection algorithm for DR screening purpose, which detected 90.1% of diabetics and 81.3% of those without DR [146]. Their system showed both sensitivity and a specificity of 93.1% and 71.6%.

## 3. Retinal Landmarks Detection

A standard CAD screening system needs to extract the retinal landmarks such as fovea, ONH, and vessels, and additionally lesions such as MAs, HEs, EXs, CWS, and NVs. Several studies have worked on various detection and segmentation approaches to identify these anatomical regions. Here, we briefly discuss them.

*3.1. Localization and Segmentation of Optic Nerve Head*

The ONH can reveal critical eye disease like glaucoma [147]. The shape of the ONH appears as semicircle or ellipse in the retinal image. The ONH is also applied as a reference landmark to identify other retinal regions and structures such as the macula and fovea [87, 148, 149, 150]. Moreover, the ONH localization assists to determine central retinal arteries and veins [87, 151]. Many vessel tracking algorithms use the ONH center as the starting point, since the vessels diverge from the ONH center. The ONH localization is carried out by detecting its center or by drawing a circle around the ONH region [152-156]. ONH segmentation is introduced by specifying the boundary of ONH. Thus, identification and extraction of ONH is highly important [135, 147, 157-161].

The localization of the ONH is serious to differentiate it from the other lesions. The ONH identification approaches, as explained in literature mainly estimate its center or surround it by a circular mask. In both cases, the localization is quite an elaborate task because of the presence of thick vessels, inaccurate boundaries, and lesions such as EXs [135, 147]. The methods according to intensity changes are easy to use, fast and robust for images with no symptoms. These methods may lose when the ONH is concealed by blood vessels and lesions such as EXs, CWS, or other same intensity artifacts [162].

### 3.1.1. Detection of Optic Nerve Head Center

The ONH color and shape can be utilized to detect it. Normally, in some studies the ONH is localized by its relations to bright pixel clusters. Sinthanayothin et al. [163] used a window size 80 × 80 pixel as the size of a typical ONH and localized the ONH by is observing the maximum change in its intensity to its neighboring pixels. They chose the point with highest pixel variance to detect the center of the ONH and reported a sensitivity and specificity of 99.1%. Lowel et al. [87] developed correlation filter to identify the ONH in presence of EXs and other artifacts and correctly detect it in 83% of cases. Hoover et al. [88] presented an approach using the concept of fuzzy approach to locate the ONH. Their approach obtained 89% correctly ONH detection. Walter et al. [164] introduced a method to detect ONH based on its illumination, distance, and watershed transformation in different color space, HSL (Hue, Saturation, Luminance), with accuracy of more than 85%. Tavakoli et al. applied three different vessel segmentation algorithms, Laplacian-of-Gaussian, Canny, and Matched filter edge detectors to detect the ONH in both the images without DR signs and in the presence of DR [165].

The Hough and similarly the Radon transform were applied by many authors to identify the ONH [53, 87, 166-168]. The method is based on integral transformation, or mapping the values of the line integral of an images pixels into another domain. Ultimately, the ONH was identified by effective projection profiles in Radon space and the boundary parts with maximum vote [53].

Principal Component Analysis (PCA) was another method for identifying the ONH ([163, 169, 170]). In fact, this method turns the correlated variable into uncorrelated ones which is known as principle parts [1]. The first principal component demonstrates the maximum information exist in the image. Li and Chutatape [170] detected the ONH using the concept of PCA with accuracy of 99%. Another method that applied the convergence of vessels to localize the ONH was proposed by Foracchia et al. [171]. Youssif et al. [172] applied matching the expected directional structure of the vessels for ONH detection. Their method started by normalizing luminosity and contrast throughout the image using illumination equalization and adaptive histogram equalization methods respectively. Lu and Lim [173] utilized a different approach to place the ONH based on its bright appearance in color retinal image. They used a series of concentric lines with various directions and assessed the image variation along the different directions. Figure 3 shows the retinal images of these categories and related signs.

*3.1.2. Segmentation of Optic Nerve Head*

In case of segmentation methods category, the brightness was applied by Li and Chutatape [174, 175] to detect the ONH candidates for their model-based technique.

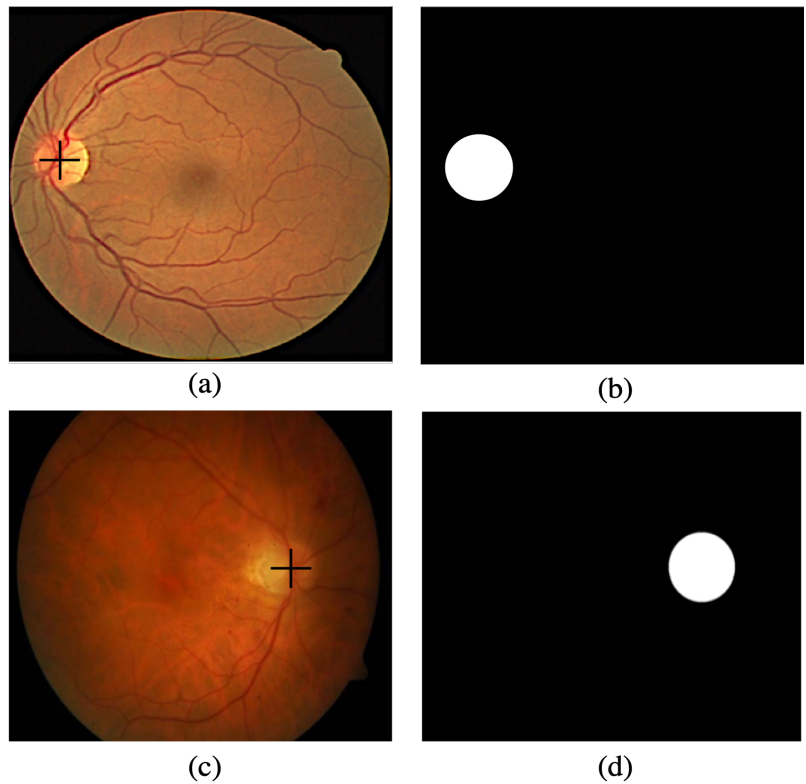

Figure 3: **ONH detection.** (a, c) Retinal images from MUMS-DB and detected center of the ONH; (b, d) ONH mask segmentation.

They established an active shape model, which consisted of building a model with training cases and iteratively matching the landmark points on the disc edges and main vessels inside the disc [170]. A model-based approach was introduced effectively based on active contour model by Osareh et al. [176, 177] to approximately detect the ONH. Lowell et al. carefully created an ONH template, which was then correlated to the intensity component of the retinal image using the full Pearson-R correlation [87]. Another model-based algorithm was introduced by Xu et al. [178]. They applied the deformable contour model technique that included ONH margin. Their method used a clustering-based classification of contour points and customized contour evolution step integrated in original snake formulation. The approach introduced by Wong et al. [179] was based on the level-set method followed by ellipse matching to smooth the boundary of the ONH. They proposed: 1) the ONH location according to the previously obtained location by means of histogram analysis and initial contour definition, and 2) subsequently, a modified version of the conventional levelset method was used for ONH boundary extraction from the red channel. This contour was finally matched by an ellipse.

On the other hand, Lu [86] introduced a circular transformation to take ONH and its boundary simultaneously. Lalonde et al. presented Hausdorff-based template matching together with pyramidal decomposition and confidence assignment [180]. Pyramidal decomposition acts mainly for detection of large areas with bright pixels, which probable existence of the ONH, but it can be disturbed by large areas of bright pixels that may be near the images borders due to non-uniform illumination. In order to solve this problem, Frank ter Haar [181] used illumination equalization in the green channel of the image, and then a resolution pyramid using a simple Haar-based discrete wavelet transform was formed. Besides, he applied the Hough transform and proposed two methods using a vascular branch constructed from a binary vessel image. Last technique of Frank ter Haar was based on fitting the vessel orientations on a directional model [181]. The ONH identification methods not only apply the characteristics of the ONH, but also extract the location and orientation of vessels [182, 183, 184, 185]. Niemeijer et al. [185] presented the use of local vessel geometry and image intensity features to find the correct positions in the image. A kNN regressor was used to accomplish the integration. Tobin et al. [182] used an algorithm that highly depended on vessel-related the ONH characteristics. They applied a Bayesian classifier to categorized each pixel in images as with/without ONH. Abramoff and Niemeijer [186] applied same properties for the ONH detection, and then a kNN regression was used to estimate the ONH location. The approach developed by Abramoff et al. [187] according to a pixel classification using the feature analysis and nearest neighbor technique. The final results of their algorithm came with the categorization of pixels to a group that belonged to the ONH and non-ONH. In a recent study, Hsiao et al. placed ONH by illumination correction operation, and contour segmentation that was completed by a supervised gradient vector flow snake [188]. To identify ONH location candidates, first template matching was used by Yu et al. Next, they applied vessel features on the ONH to place the location of it. They combined region and local gradient information to the segmentation of the ONH boundary and used Morphological filtering to remove retinal vessels [189].

Moreover, some approaches made use of the fact that all major vessels diverge from the ONH [171, 172, 181, 190]. Most of the studies identified the ONH respect to its shape and color. Other algorithms recognized the ONH according to tracking vessels toward their origin.

*Table 5: Summarize of some methods and their results of ONH detection*

| Authors | ONH detection method | DRIVE | STARE |
|---|---|---|---|
| Sinthabayothin *et al*. [163] | Highest average variation | 99.1% | 42% |
| Walter and Klein [164] | Largest brightest connected object | 100% | 58% |
| Li and chutatape [174,175] | Brightness guided, model base | 99% | - |
| Osareh *et al*. [176,177] | Averaged ONH model based | 100% | 58% |
| Lowell *et al*. [87] | ONH Laplacian of Gaussian template | 99% | - |
| Lalonde *et al*. [157] | Hausdorff based with pyramidal decomposition and confidence assignment | 100% | 71.6% |
| Frank ter Haar [181] | Resolution pyramid using Haar based wavelet transform | 89% | 70.4% |
| Frank ter Haar [181] | Hough used only to pixels on or close to the retinal vasculature | 96.3% | 71.6% |
| Frank ter Haar [181] | Pyramidal decomposition of both the vascular and green plan | 93.2% | 54.3% |
| Frank ter Haar [181] | Hough and fuzzy convergence | 97.4% | 65.4% |
| Tobin *et al*. [182] | Vasculature related ONH features and a Bayesian classifier | 81% | - |
| Abramoff & Niemeijer [185,186] | Vasculature related ONH properties and a kNN regression | 99.9% | - |
| Youssif *et al*.[172] | matching the expected directional pattern of the retinal blood vessels | 100% | 98.8% |
| Hsiao *et al*. [188] | illumination correction, and contour segmentation is completed by a supervised gradient vector flow snake | 100% | 90% |
| Tavakoli et al. [165,168] | Different vessel segmentation methods | 90% | 85% |
| Pourreza et al. [53] | Radon Transform | 100% | 94% |

## 3.2. Vessel Segmentation

There is a substantial effort reported in the state of art for the segmentation of retinal vessels in fundus images [56, 91, 94, 95, 97, 191-196]. Overall, the vessel segmentation approaches are classified in to five categories: *tracking, mathematical morphology, matching filter, model-based thresholding, and supervised classification algorithms* [1]. Table 6 summarizes all these methods and below, we briefly are discussed below.

*Table 6: Some vessel segmentation methods*

| Authors | Method | Classifier & Databases | Evaluation |
|---|---|---|---|
| Xu and Luo [92] | Sobel filter for ONH, Adaptive threshold | Thick & thin vessels, SVM, DRIVE | Acc.= 93.36% Sen. = 85.57% |
| Hassan et al. [197] | Hidden Markov Model | Vessels, DRIVE | Acc.= 95.7% Sen.= 81.0% Sp. = 97.0% |
| Dash and Bhoi [89] | Binary difference and enhanced image | DERIVE, CHASEDB1 | Acc.= 95.5%, |
| Nguyen et al. [93] | Linear combination of line responses | SVM, STARE, DRIVE, REVIEW | Acc.= 93.24%, 94.07%, |
| Fraz et al. [56, 95] | Mathematical morphology, aggregation | DRIVE, STARE MESSIDOR | Acc.= 95.52% Sp. = 97.23% |
| Marin et al. [96] | Grey level features | ANN, DRIVE, STARE | Acc.= 95.26%, Sen.= 69.44 %sp.= 98.19% |
| Aslani and Sarnel [198] | Hybrid features | DRIVE, STARE | Acc.= 95.13%, Sen.= 77.96% sp. =97.17% |
| Lupascu and Trucco [199] | Ada Boost | Ada Boost classifier, DRIVE | Acc.= 95.69% Sen.= 55.74% sp.= 99.36% |
| Rodrigues et al. [160] | Histogram analysis, Wavelet hessian based and Gaussian filter, | DRIVE, HRF, | Acc.= 94.65%, |
| Tavakoli et al. [101] | Radon transform | MUMS-DB | Acc.=90%, Sen.=85% |
| Zhang et al. [94] | kNN | DRIVE | Acc.= 95.05% Sen.= 78.12% Sp.= 96.62% |
| Zhu et al. [90] | Vessel profile, Gaussian, Morphology | ELM, DRIVE, RIS | Acc.= 96.07, Sen.= 60.18, Sp.= 98.68% |
| Neto et al. [91] | Binary morphological reconstruction | DRIVE, STARE | Acc.= 86.16%, 87.87% Sen. = 79.42, 76.96% Sp.= 95.37%, 96.31% |
| Vega et al. [200] | Lattice NN and dendritic processing | ANN | Acc.= 94.83%, Sen.= 96.71%, Sp.= 70.19% |
| Waheed et al. [201] | Connected component, Binary threshold, | Localized Fisher Discriminant Analysis, DRIVE, STARE | Acc.= 95.81%, 96.16%, |
| Zilly et al. [202] | Ada Boost | MESSIDOR, DRISHTIGS | - |

A comprehensive review on existing methods in retinal vessel segmentation and available public datasets are presented in [52, 33]. In general, vessel segmentation algorithms can be classified into two broad categories: unsupervised and supervised methods [203]. A comparative survey of these two classes has been presented in [102]. Unsupervised methods can be further classified into techniques based on matched filtering [47, 99], morphological processing [204], vessel tracking, multi-scale analysis [205, 206], line detectors [93] and modelbased algorithms [98, 207]. Supervised segmentation methods are based on pixel classification such as the k-nearest neighbors (kNN) [49], Gaussian mixture models [107], SVM [208], NN [208], decision trees [195], and AdaBoost [199, 96]. They utilize ground truth data for the classification of vessels, based on given features.

The principle of matched filter detection in unsupervised methods was proposed in [47]. With this technique, the authors used a 2D Gaussian-shaped template in all directions to search for vessel components. The kernel was rotated through a range of angles (usually 8 or 12) to detect vessels with different orientations. The resulting image is thresholded to produce a binary representation of the retinal vasculature. Another matched filtering approach used for retinal vessel segmentation [209]. The application of local thresholding strategies, edge detection, matched-filtering and region growing was developed in [209]. Chakraborti et al. [99] applied an unsupervised segmentation approach that combined a "vesselness" filter and matched filter using an orientation histogram. Vermeer et al. [98] achieved vessel detection by thresholding, after convolving the image with a 2D Laplacian kernel. A general framework of adaptive local thresholding based on combination of applying a multithreshold scheme and classification procedure to prove all binary result objects were applied by Jiang et al. [210]. Zana et al. [106] proposed a morphology-based approach using an algorithm that combined morphological filters and cross-curvature evaluation to segment vessel-like patterns in retinal images. Other mathematical morphology approaches are described in [204, 105, 211]. Heneghan et al. [211] combined morphological transformations with curvature information and matched filtering for centerline detection. In [100] vessels are extracted using multiple structure elements followed by connected component analysis. Martinez-Perez et al. [205] introduced an automated algorithm for retinal images based on a multi-scale feature detection. The local maxima of the gradient magnitude and the maximum principal curvature of the Hessian tensor were used in a multi-pass region growing process. This procedure accurately segmented the retinal vasculatures by using both feature and spatial information. Zhang et al. [206] employed innovative rotating multi-scale second-order Gaussian derivative filters which are referred to as orientation scores for the enhancement and segmentation of blood vessels. Roychowdhury et al. proposed an iterative unsupervised retinal vessel segmentation algorithm that used an adaptive threshold and a region growing method with a stopping criterion to terminate the iteration [212].

Nguyen et al. proposed a vessel detection method based on line detection [93]. This approach was based on this fact that changing the length of a line detector makes line detectors with variety of scales. The final vessel segmentation results were achieved by combining line responses at varying scales. The combination of shifted filter responses for detection of bar-shaped structures in retinal images was presented by Azzopardi et al. [104]. Their approach was rotation invariant, where the selection of the orientations was chosen from given the structures similar to vessels which suffered from difficult crossing cases. Lam and Yan [207] used the Laplacian operator to segment blood vessels and detect centerlines from the normalized gradient vector field, pruning noisy objects. Vessel continuity employs measures of width and orientation, iteratively calculated in a local region near the current point, in order to track along the length of a vessel [100, 103]. Tavakoli et al. [101] presented the combination of Radon transformation and multi-overlapping windows to segment the retinal vessels. Since their approach relied on the concept of Radon, which is integral based, they suppressed the intrinsic noise of images by using this transformation.

Supervised approaches use a pixel classification method, referred to as a primitive-based approach by Staal et al. [213]. This approach was based on the detection of ridges, used as primitives for introducing linear segments, called line elements. Sinthanayothin et al. [163] identified retinal vessels using a NN whose inputs were deduced from PCA of the image and edge extraction of the first principal component. A NN and backpropagation multilayer method was proposed by Gardner et al. [123] for vessel tree detection following histogram equalization. The kNN classifier was implemented by Niemeijer et al. [49]. In their algorithm, from the green plane of the RGB image, a vector of 31-component pixel feature was created with the Gaussian and its derivatives up to order 2 at 5 different scales. Soares et al. [107] applied a Gaussian mixture model Bayesian classifier. By using the Gabor wavelet transform, multiscale analysis was performed on the image. Ricci and Perfetti [208] utilized a SVM for classifying pixels as vessel or non-vessel. In their algorithm, along with the gray-level of the target pixel they applied two orthogonal line detectors to create the feature vector. One of the interesting supervised methods was implemented by Fraz et al. [56] which was an ensemble classification system of boosted and bagged decision trees. The drawback of supervised classifications is its necessity for a sufficient number of manually annotated training image set. Moreover, they are not easy to generalize the trained models to meet the needs of different datasets. More recent studies have successfully applied the concept of deep learning to the segmentation of the retinal vasculature [90, 214-223].

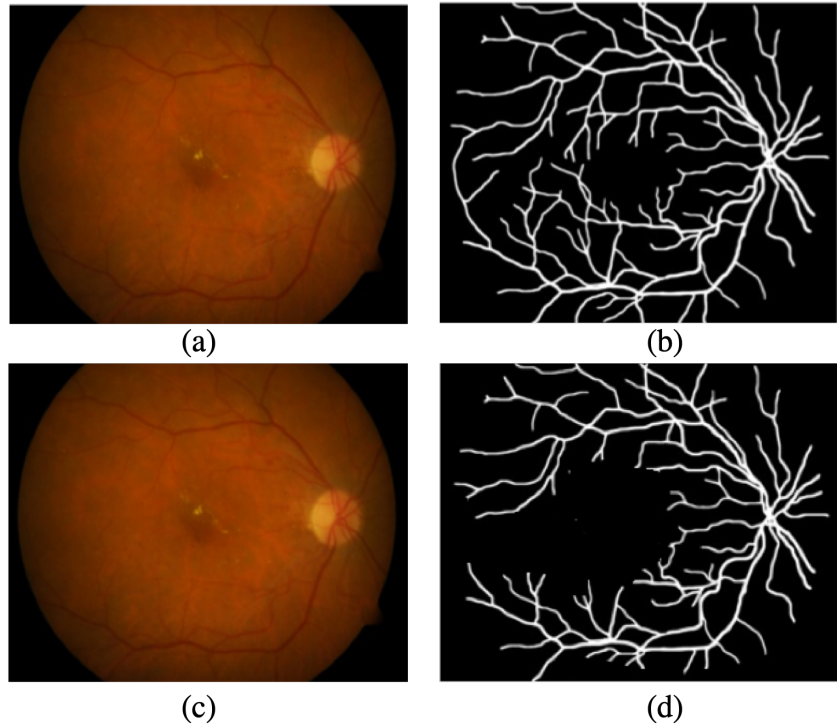

Figure 4: **Vessel segmentation.** (a, c) Retinal images from MUMS-DB; (b, d) Vessel segmentation results.

*3.3. Localization of the Fovea*

Fovea is small region approximately at the center of retina and responsible for central and high resolution vision. It is placed around 45 times of ONH radius to the temporal direction of ONH [163, 170, 185, 224, 225, 226, 152, 227, 228, 229]. The visual cells placed in the fovea are closely packed and cause the highest sharpness of vision [230]. No blood vessels passing through the fovea to obstruct the crossing of light considerable the foveal cells [231]. The fovea detection approaches are divided in to two classes: hybrid and constraints based approaches [1].

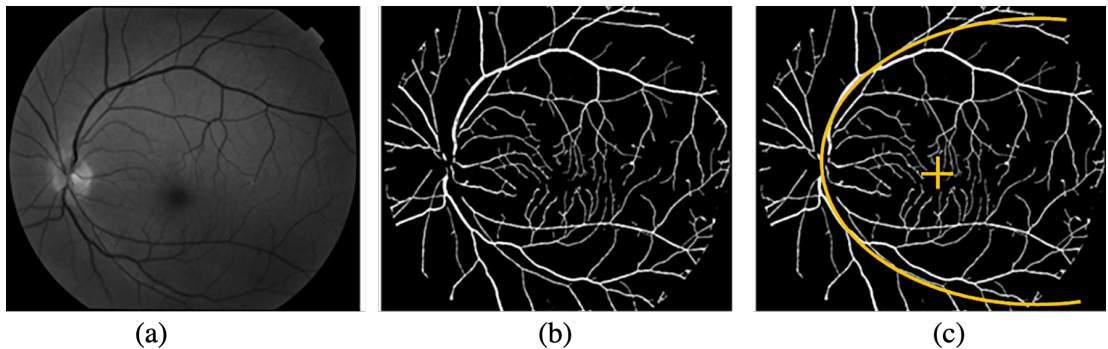

Figure 5: **Fovea detection.** (a) Retinal images from MUMS-DB; (b) Vessel segmentation results; (c) Detected fovea.

### 3.3.1. Hybrid-based method

Sinthanayothin et al. [163] presented a method for fovea localization using intensity correlation. At the beginning, fovea was behaved with intensity correlation and from HSI transformation the peak was selected. The selected peak located a dark area and was considered as fovea. The disadvantageous of their method was when the fovea was not at the center it failed. In another method, using location of fovea, a polar retina coordinate system was established. Here, the fovea accurately was located using its appearance and geometry according to other landmarks [170]. Niemeijer et al. [185] proposed a method to identify the fovea based on single point distribution model. The method included all signs including global and regional, to identify this model. These signs were extracted from the related vessel position and its width. Tavakoli et al. [148] proposed an approach based on vessel segmentation and parabola fitting to identify the fovea. Their algorithm was included of 4 steps: multi-overlapping windows, local Radon transform, vessel validation, and parabolic fitting.

### 3.3.2. Positional constraints-based

The positional restrictions ( [163, 170, 184, 185, 224, 225, 226]) were applied to localize the fovea and determine its position. The fovea identification was obtained using the location of other retinal regions. The main part of this approach [182] specifies the horizontal retinal ridge that is passing across the ONH and fovea, which divides the upper and lower retinal regions. Further, the region of fovea was extracted using fixed distance of 5 ONH radius from the ONH center.

*Table 7: Details of some of fovea detection methods.*

| Author | Approach | No. of images | Accuracy |
|---|---|---|---|
| Sinthabayothin et al. [163] | NN and template-based | 112 | 80.4% |
| Li and Chutatape [175] | Modified active shape model | 35 | 100% |
| Tavakoli et al. [148] | Fitting the main vessels with parabola function | 140 | 90% |
| Tobin et al. [182] | ONH and fovea localization | 345 | 92.5% |
| Seckar et al. [232] | Hough transform and morphological function | 44 | 100% |
| Fleming et al. [184] | Vessel segmentation with semiellipse fitting | 1056 | 96.6% |
| Niemeijer et al. [185] | Point distribution model and kNN | 500 | 96.8% |

| Welfer et al. [233] | ROI Sselection and morphological operator | 126 | 96.1% |

## 4. Detection of retinal lesions

An acceptable CAD system requires to identify lesions such as MAs, HEs, EXs, CWS, and NVs as perfect as possible to finally result in distinguishing between DR and no DR image. Several studies have been worked on various detection and segmentation approaches to detect these lesions. Here, we briefly have discussed them.

*4.1. Microaneurysms and Haemorrhages Detection:*

The most reliable method for MA detection is from fluorescein angiography retinal images [19]. However, the fluorescein angiography images are costly, invasive and not suitable for everyone [22]. Moreover, this approach is time-consuming. Introductory global image processing approaches were applied for automated detection. There are a number of methods for the detection of both MAs and HEs. These methods can be generally categorized into two general groups including unsupervised and supervised approaches [17, 22, 108, 109, 111, 112, 227, 234-248]. Many studies have searched both MAs and HEs detection from the retinal images by suing proper thresholding after masking the landmarks such as ONH, fovea, and vessel. After the thresholding, mathematical morphology was used to distinguish MAs and other structures, such as bifurcations and small vessels. From unsupervised viewpoint, the MAs detection approaches are classified in to different classes: morphology, region growing, wavelet, and hybrid approaches [1]. Among these groups, the HEs detection is mostly based on mathematical morphology combine with pixel classification. In this section, we briefly explain the related studies.

Walter et al. [22] presented a 4-part method for MAs detection. At the beginning, the preprocessing step used green channel images. The top-hat transformation and global thresholding were used to identify MAs. Next, fifteen characteristics were used to differentiate true MAs from false ones using kNN, Gaussian and kernel density classifiers. Hatanaka et al. [249] proposed a brightness correction approach for automated detection of HEs. The correction was applied in Hue Saturation Value (HSV) space.

In the morphological approach, different adjustments often increase the accuracy of MAs detection [17, 241]. In spite of this type of hand-crafted method, typically is fast, easy, and user friendly but the ability of it is constrained by its builder. In better words, some main hidden structures and uncovered patterns might be ignored by the builder and cause false detection [111]. The morphological operations such as closing [22], or using top-hat transformation [19] has been applied for MAs detection due to their relatively uniform circular shape and limited size range.

However, besides the above issue, most of the mathematical morphology methods mainly depend on the choosing of optimal structuring elements; changing their size and shape decreases the efficiency of these algorithms.

Fleming et al. [17] developed contrast normalization, watershed and region growing approaches to detect MAs in retinal images. Moreover, they applied a 2D Gaussian-aftermedian filtering to preprocess the image. To detect MAs they used the kNN classification using nine characteristics from extracted candidates. Sinthanayothin et al. [21] identified MAs by applying region growing and adaptive intensity thresholding, combined with a moat operator. Quellec et al. [108] used wavelet transform algorithm for MAs detection by matching wavelet sub-bands with the lesion template.

Tavakoli et al. [19] employed an approach based on top-hat, mathematical morphology and Radon transform, applied locally in multi-overlapping windows to detect MAs. Niemeijer et al. [234] introduced a hybrid red lesion detection method using ideas from Spencer et al. [250] and Frame et al. [251]. At the beginning, using pixel classification the red lesion candidates were identified [250]. The MAs and HEs were detected after masking the vessels. Usher et al. [132] and Gardner et al. [123] applied NN to locate MAs. Initially, they used local adaptive contrast enhancement to preprocess the image. After that landmarks (ONH and vessel segmentation), and red lesions (MAs and HEs) were identified using region growing and adaptive intensity thresholding combined with moat operation [21]. Zhang et al. [252] described a method using Multi-Scale Correlation Filtering and threshold values within angiographic retinal image for MA detection. They used a moving Gaussian kernels to extract the correlation coefficient for pixels. Larsen et al. [146] introduced an automatic MAs and HEs detection approach using classification algorithm called RetinaLyze using shape and size of the lesions.

Fleming et al. [133] applied multiscale and morphological operators to place HEs. For the preprocessing, they corrected the variations in intensity and contrast by applying median filter and histogram equalization. In another study [17], they proposed a region growing algorithm to identify large lesion regions. In their approach, they had difficulty differentiating HEs and MAs. Zhang and Chutatape [253] introduced an approach to detect HEs using SVM. In addition, they applied PCA to detect the features. These features were used as the inputs of SVM to extract HEs. Hipwell et al. [254] proposed an automated MA detection in red-free retinal images. To remove variation in the illumination, a shading correction was applied to the image.

Another approach in contrast with linear landmarks like vessels, which are directional, is to utilize template matching filters with multiscale Gaussian kernels [108, 252, 236]. According to statistical results, the intensity distribution of MAs is matched to a Gaussian distribution [108] because MAs display Gaussian profiles in all projections. Therefore, template matching filter highly increased the accuracy of MA detection. Using this idea, Quellec et al. proposed a wavelet transform method for MA detection. In their algorithm, they detected the MAs using a local template matching filter in the wavelet domain [108]. Zhang et al. [252]used a multi-scale correlation coefficient based approach. In their method, they used a nonlinear filter with five Gaussian kernels at various standard deviations to detect MA candidates. Ram et al. [240] described a feature based method that rejects specific classes of clutter while accepting a higher number of true MAs. The potential MAs, obtained after the final step, are labeled a grade that was based on their shape similarity to true MAs. Same as the morphological approach, this type of algorithms is still limited by ignorance of hidden and unnoticeable structures. Moreover, a lot of parameters required to be assigned empirically.

The brief explanation of the MAs and HEs detection algorithms is introduced in Table 8.

*Table 8: Details of some of MA and HE detection methods*

| Authors | Methods | Performance |
|---|---|---|
| **MA detection** | | |
| Tavakoli et al. (2013) [19] | Radon transform | Sen. = 92%, Sp. = 90% |
| Hipwell et al.(2000) [254] | Matched filter, region growing | Sen. =81%, Sp. = 93% |
| Sinthanayothin et al. (2002) [21] | Moat operator | Sen. =77.50%, Sp. = 88.70% |
| Larsen et al. (2003) [146] | Size and shape | Sen. =71.4% |
| Usher et al. (2004) [132] | Moat operator | Sen. =95.10%, Sp. = 46.30% |
| Niemeijer (2005) [234] | Pixel classification using kNN | Sen. =100%, Sp. = 87% |
| Fleming et al. (2006) [17] | Contrast normalization, watershed region growing | Sen. =85.40%, Sp. = 83.10% |
| Walter et al. (2007) [22] | Gaussian filtering, top-hat | Sen. =88.5% |
| Quellec et al. (2008) [108] | Optimal wavelet transform | Sen. = 89.62%, Sp. = 89.50% |
| Zhang et al. (2010) [252] | Multi-scale correlation filtering and dynamic thresholding | Sen. =71.30% |
| Sanchez (2011) [73] | | Sen. =91% |
| Antal and Hajdu (2012) [255] | Ensemble-based | AUC=0.90 |
| Lazar and Hajdu (2013) [236] | Cross section features | - |
| Pereira (2014) [113] | multi-agent system approach | Sen. = 87% |
| Seoud (2016) [23] | Dynamic Shape Features | Sen. = 84.4% |
| Srivastava (2017) [110] | Filters for feature extraction | - |
| **HE detection** | | |
| Gardner (1996) [123] | NN and Statistical threshold tuning | Sen. = 73.80% |
| Zhang and Chutatape (2005) [253] | PCA and SVM | Sen. = 89.10% |
| Fleming et al. (2008) [133] | Multi-scale, morphology and SVM | Sen. = 98.60%, Sp. = 95.50% |
| Hatanaka et al. (2008) [249] | Brightness correction and thresholding | Sen. = 80%, Sp. = 88% |

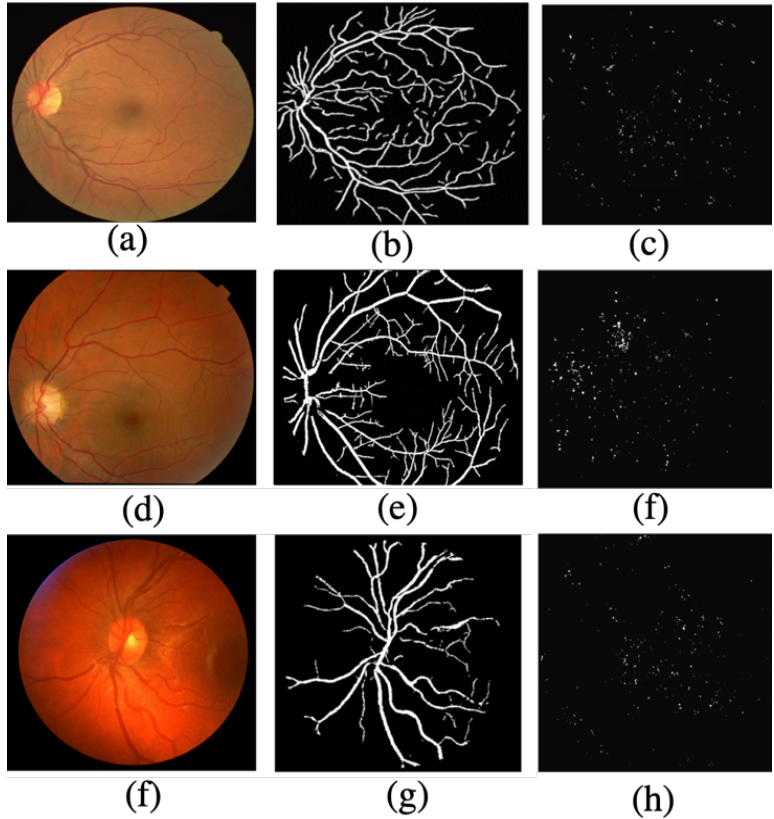

Figure 6: **MA detection.** (a, d, f) Retinal images from different datasets; (b, e, g) Vessel segmentation and masking results; (c, f, h) Detected MAs.

*4.2. Exudates Segmentation*

Here, we briefly review the studies of EXs segmentation. EXs are one of the main signs of DR, because they cause retinal edema and consequently vision loss and blindness. The approaches for identification of EXs [256, 257, 258, 114] can be categorized into four classes: pixel clustering, morphology, thresholding/region growing, and classification-based approaches [1]. Here, we concisely introduce these methods.

Several studies [120, 259, 260] have used Fuzzy C-Means cluster-based method for EXs detection. This method suggests pixels with various classes and changing degrees of membership for segmentation purpose [261, 262]. Osareh et al. [259] first by correcting color and contrast worked on image preprocessing. The preprocessed images were used for segmentation step using Fuzzy C-Means. The genetic algorithm was applied to choose characteristics from Fuzzy C-Means clustered regions [260]. The optimal characteristics were then used in the NN to detect EXs and non-EXs regions. This approach cops the differing retinal color. In another study, the EXs were segmented from low quality images using Fuzzy C-Means [120]. The same authors presented EXs detection using Bayesian classifier [120]. Hsu et al. [69] described an algorithm of EXs segmentation using cluster-based approach. Then, cluster-based method was utilized to classify the abnormal pixels. The restrictions of the above methods is to know considerable clusters which contribute to the EXs detection.

Morphological [263, 264, 121] and thresholding/region growing methods [21, 136] were applied in many studies about EXs detection. Sopharak et al. [263] presented an EXs segmentation approach for low image quality by applying adjusted morphology. Walter et al. [264] detected EXs using gray level changes of green channel. After first step detection, their boundaries were determined applying a morphological reconstruction approach. Their method did not differentiate EXs from CWS. Sanchez et al. [121] used same approach as Walter et al. [264] used in preprocessed the green channel and histogram of the image was modelled using a Gaussian Mixture Model. Furthermore, a thresholding method was utilized to detect the EXs using Gaussian Mixture Model. Sinthanayothin et al. [21] developed a region growing-based method by selecting some threshold values in the gray scale images. To improve the quality of image, adaptive local contrast enhancement was applied. This algorithm located other retinal regions and landmarks such as the ONH, retinal vessels and fovea. Fleming et al. [184] presented a multi-scale morphology algorithm to localize EXs. They used Gaussian and median filter to preprocess the green channel images for correcting the contrast variations. A watershed region growing algorithm was used to separate the EXs from other landmarks. Finally, They sued SVM to classify regions into EXs and non-EXs. Welfer et al. [265] presented an EXs detection algorithm based on Morphology operator. They applied a set of morphological functions to detect EXs. Mookiah et al. [136] introduced EXs detection by applying color, shape and morphological processing. Their method assumed EXs to have higher contrast than the ONH.

Several studies [123, 234, 266] utilized machine learning approaches to categorize EXs and non-EXs pixels. Gardner et al. [123] utilized NN to classify EXs applying gray scale of retinal images. Sanchez et al. [266] segmented EXs according to Fishers Linear Discriminant Analysis. They used color normalization and contrast enhancement. Niemeijer et al. [234] introduced an automated method for EXs detection using classification based on machine learning idea. The pixels with high probability were classified into EX class from non-EXs.

5. **Artificial Intelligence in Diabetic Retinopathy**

In general, understanding and controlling DR or other retinal diseases has become extensively more complicated because of the large number of images and diagnoses. In fact, processing all these patient examinations seems as a big data challenge [267]. Clearly, the new era of diagnosis and clinical data mining immediately needs intelligent systems to control them sufficiently, safely and efficiently.

Recently, through the advancement of supervised learning ideas, the recent application of artificial intelligence (AI) approaches for DR detection and its related lesions have significantly increased in the research community. AI has already revealed proof-of-concept in medical science such as radiology and pathology, which like ophthalmology relies heavily on diagnostic imaging. As its turning out, image processing is the most important application of AI in healthcare [268]. Moreover, AI is particularly suitable in assisting clinical tasks by using efficient algorithms to identify and learn features from large volumes of imaging data, and helping to reduce diagnostic errors. In addition, it can recognize specific patterns or lesions and correlate novel characteristics to obtain innovative scientific insight [269].

The most important branch of AI is machine learning (ML) [269] where the rules would be learned directly by algorithms from a series of examples, called training data, instead of being encoded by hand. ML techniques need that a set of biomarkers or characteristics to be measured directly from the training data (e.g., labeled lesions in retinal image). After that based on a training database of characteristics with known labels, i.e. hand crafted features, a classifier learns to identify the correct label from the newly seen characteristics. Once a few strong classifiers have been established, the effectiveness of such ML models mostly relies on the differentiative power of the chosen characteristics which underpin the classifier performance. There are different reviews that have summarized ML approaches [1, 15, 74, 270-273]. Both Mookiah et al. [1] and Mansour [74] classified DR diagnosis methods in respect to affiliated methods, such as DR lesion tracking, mathematical morphology, cluster-based, matched filter, and hybrid methods. Faust et al. [15] surveyed studies that detect lesion characteristics from retinal images, such as the retinal vessels, MAs, HEs, and EXs. Joshi and Karule [270] summarized the first researches on EXs segmentation. Almotiri et al. [271] came up with a survey of methods to detect the vessels. Almazroa et al. [272] and Thakur and Juneja [273] studied several approaches for ONH detection.

Recently, the accessibility of large databases and the immense computing power suggested by advanced graphics processing units have research on Deep Learning (DL) based methods, which have demonstrated magnificent efficiency in all medical sciences. Many DL-based approaches have been established for variety of works to analyze images to establish automated CAD systems for detection of DR. The relation between AL, ML, and DL has been shown in the figure below:

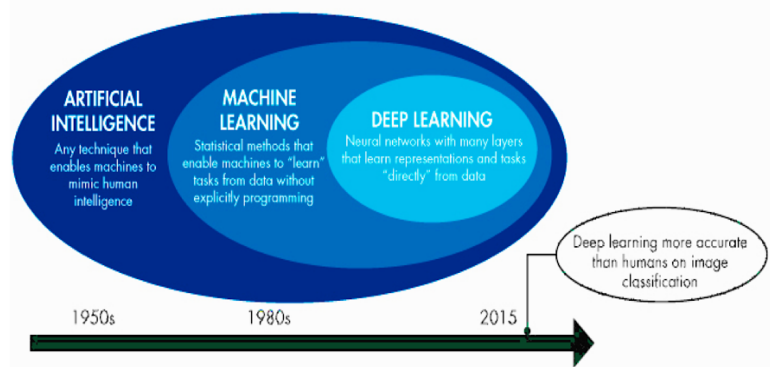

Figure 7: **AI, ML, and DL.** Relation between artificial intelligence, machine learning and deep learning.

## 5.1. Overview of Deep Learning

There are multiple DL-based algorithm that have been presented. Some frequently utilized deep architectures for the DR detection purpose include convolutional neural networks (CNNs), autoencoders (AEs), recurrent neural networks (RNNs) and deep belief networks (DBNs). A complete overview of these architectures is found in [274]. There are several DL-based architectures related to DR detection [275-285]. Based on the clinical significance of automated DR detection, we group application of DL into three broad classes: (i) vessels segmentation, (ii) ONH localization and segmentation, and (iii) retinal lesions detection and classification. In the below sub-sections, we briefly introduce the literatures about DL-based algorithms for these tasks. The brief summary of the DL methods in DR detection systems is presented in Table 9.

### 5.1.1. Retinal Blood Vessel Segmentation

Many of the retinal vessel segmentation approaches are based on CNN architectures [217, 218, 286-290].

Maji et al. [286] proposed 12 CNN models for this purpose. Each CNN model included three convolutional layers and two fully connected layers. Liskowski and Krawiec [218] introduced a supervised pixel-wise algorithm using deep CNN, which was trained by preprocessed retinal images, and augmented using geometric transformations. Tan et al. [217] employed a 7-layer CNN model to segment blood vessels, ONH, and fovea.

Some methods use AEs in segmentation of retinal vessels [291, 292]. Roy and Sheet [291] presented an AE-based deep NN model that used the domain adaptation method for training the model. AE-based deep NN included of two hidden layers, which were trained using the source domain databases, using an autoencoding and supervised mechanism. The algorithm introduced by Maji et al. [286] employed a hybrid DL, which consisted of unsupervised stacked denoising AEs with total 500 layers, to segment vessels in retinal images.

*5.1.2. Optic Nerve Head Detection*

As we mentioned at the beginning of this chapter, ONH identification can improve DR detection and classification. In fact, the ONH brightness can cause uncertainty for detecting of other bright lesions such as EXs. There are number of studies that work on both ONH detection and segmentation using concept of DL [202, 293-296].

Lim et al. [293] presented a nine-layer CNN model for ONH segmentation. Their algorithm involved different steps: detecting the region around the ONH, enhancement, classification at pixel-level using a CNN model to predict the ONH. Tan et al. [217] identified the ONH and vessels jointly using pixel patch-based CNN. Zilly et al. [202] presented an ONH segmentation algorithm based on a multiscale two-layer CNN model. At first, the area around the ONH was cropped, and down-sampled. After that, the area was analyzed by entropy filtering to determine the most distinguished points and was passed to the CNN model for final segmentation. Zhang et al. [294] applied a faster CNN to localize the ONH. After identifying the ONH, the thick vessels inside its area were eliminated by using a Hessian matrix, and its boundaries segmented.

*5.1.3. Microaneurysms and Hemorrhages detection*

Many DL-based approaches have been introduced for detection and classification of different types of DR lesions such as MAs, and HEs [2, 6, 16, 112, 115, 275, 276, 297-301].

Haloi [297] introduced a nine-layer CNN model to group each pixel as MA or non-MA and grading the DR. Each pixel is categorized by taking a window around the candidate and passing it to the CNN model. For training, they applied a data augmentation method to create six windows around each pixel. van Grinsven et al. [112] worked on detecting HEs. Their main contribution of was to overcome the over-represented normal samples created for training a CNN model. To figure out this issue, they employed a dynamic selective sampling idea that selects informative training samples. Their CNN was trained using a dynamic selective sampling strategy as well.

Gondal et al. [302] and Quellec et al. [277] also detected HEs and small red dots. Another similar approach was introduced by Orlando et al. [106]. In their approach, they first extracted red candidates using morphological operations. Next, they extracted CNN and hand-engineered characteristics such as intensity and shape from each candidate to introduce a probability map, which was utilized to make both red lesion and image level decision. Shan and Li [299] applied the stacked sparse AEs to identify MAs. A patch was passed to stacked sparse AEs, which extracted characteristics, and the Softmax classifier labeled it as a MA or non-MA patch. Tavakoli et al. employed a CNN approach to detect MAs. In their approach, they first extracted candidates using morphological functions. Next, they used CNN classification to classify between MAs and non-MAs. In their study, the authors compared effect of two preprocessing methods, Illumination

Equalization, and Top-hat transformation, on retinal images to detect MAs using combination of Matching based approach and CNN methods [6].

*5.1.4. Exudates Detection*

There are different studies on EX detections using DL [303, 304, 302, 277]. Prentasic and Loncaric [304] employed a CNN-based algorithm for EXs detection in retinal images. First, they detected the ONH, created a probability map for that. Then they created for both vessels and bright-boundary probability maps. At the end, they applied an 11-layer CNN model to produce an EX probability map and finally segment the EXs. Gondal et al. [302] presented an algorithm for EXs detection combine with other DR lesions based on CNN architecture [305]. To identify DR lesions including different types of EXs and HEs, the dense layers were removed from the CNN model. For detecting DR lesions such as EXs at the pixel level, Quellec et al. [277] proposed a CNN visualization methods. The authors did a modification on their CNN based on the sensitivity evaluation done by Simonyan and Zisserman [306], which helped in detecting DR and lesions by optimizing CNN predictions.

*Table 9: Details of some of MA and HE detection methods*

| Authors | Methods | Training | Type | Database | Performance |
|---|---|---|---|---|---|
| Maji et al. [286] | Patch-based ensemble of CNN | End-to-end | Vessel segmentation | DRIVE | AUC=0.9283 ACC= 94.7 |
| Liskowski and Krawiec [218] | Patch-based CNN | End-to-end | Vessel segmentation | DRIVE, STARE, CHASE | AUC=0.9790, 0.9928, 0.988 ACC=95.35, 97.29, 96.96 |
| Maninis et al. [218] | Fully CNN (FCN) | Transfer learning | Vessel segmentation | DRIVE, STARE | - |
| Wu et al. [100] | tracking/patch-based CNN/PCA as classifier | End-to-end | Vessel segmentation | DRIVE | AUC=0.9701 |
| Dasgupta and Singh [288] | Patch-based FCN | End-to-end | Vessel segmentation | DRIVE | AUC=0.974 ACC=95.33 |
| Tan et al. [217] | Patch-based CNN | End-to-end | Vessel segmentation | DRIVE, STARE | ACC=94.70, 95.45 |
| Mo and Zhang [290] | Multi-level FCN | End-to-end | Vessel segmentation | DRIVE, STARE, CHASE | AUC=0.9782, 0.9885, 0.9812 ACC=95.21, 96.76, 95.99 |
| Maji et al. [286] | Patch-based AE | Transfer learning | Vessel segmentation | DRIVE | AUC=0.9195, ACC=93.27 |
| Li et al. [292] | Patch-based AE | End-to-end | Vessel segmentation | DRIVE, STARE, CHASE | AUC=0.9738, - , 0.9716, ACC=95.27, 96.28, 95.81 |
| Lahiri et al. [291] | Patch-based AE | Transfer learning | Vessel segmentation | DRIVE | ACC=95.3 |
| Fu et al. [289] | Patch-based CNN as RNN | End-to-end | Vessel segmentation | DRIVE, STARE, CHASE | ACC=95.23, 95.85, 94.89 |
| Lim et al. [293] | CNN with exaggeration | End-to-end | ONH segmentation | MESSIDOR, SEED-DB | - |
| Tan et al. [217] | CNN | End-to-end | ONH segmentation | DRIVE | ACC=87.90 |
| Sevastopolsky [295] | Modified U-Net CNN | Transfer learning | ONH segmentation | DRION-DB | - |
| Zilly et al. [202] | Multi-scale CNN | End-to-end | ONH segmentation | DRISHTI-GS | - |
| Alghamdi et al. [296] | Cascade CNN, each model | End-to-end | ONH detection | DRIVE, DIARETDB1, MESSIDOR, STARE, | ACC=100, 98.88, 99.20, 86.71 |

| Xu et al. [178] | CNN based on deconvolution | Transfer learning | ONH detection | MESSIDOR, STARE | ACC=99.43, 89 |
|---|---|---|---|---|---|
| Haloi [297] | 9-layer CNN | End-to-end | MA detection | MESSIDOR, ROC | AUC=0.982 ACC=95.4 |
| van Grinsven et al. [112] | Patches based selective sampling | End-to-end | HE detection | Kaggle, MESSIDOR | AUC=0.917, 0.979 |
| Gondal et al. [302] | CNN model | Transfer learning | HE detection | DIARETDB1 | - |
| Quellec et al. [277] | CNN model | Transfer learning | HE detection | DIARETDB1 | AUC =0.614 |
| Orlando et al. [115] | HEF + CNN and RF classifier | End-to-end | MA-HE detection | DIARETDB1, e-ophtha, MESSIDOR | AUC=0.8932 |
| Tavakoli et al. [2,6,16] | Patches CNN | End-to-end | MA detection | DRIVE, DIARETDB1, MESSIDOR | ACC=95.97 |
| Shan and Li [299] | Patches based AE | Transfer learning | MA detection | DIARETDB | ACC=91.38 |
| Prentasic and Loncaric [304] | 11-layer CNN, | End-to-end | EX detection | DRiDB | - |
| Gondal et al. [302] | CNN model | Transfer learning | EX detection | DIARETDB1 | |
| Quellec et al. [277] | CNN model | Transfer learning | EX detection | DIARETDB1 | AUC=0.974, |

## 6. Conclusion

This chapter introduces a detailed studies of methods and their results for DR detection and classification using retinal image. The DR is a complication of diabetes that damages the retina, and causes vision loss and blindness. A robust DR screening system will remarkably decrease the workload of clinicians. The screening process brings sets of steps namely identifying and segmenting the retinal landmarks, detecting lesions, feature extraction and classification. All these steps need different methods and techniques. Although considerable accomplishments have been made in digital retinal image analysis, there are variety of challenges in the selection of suitable approaches which result in high accuracy during DR screening.

[305] Asiri N, Hussain M, Al Adel F, Alzaidi N. Deep learning based computer-aided diagnosis systems for diabetic retinopathy: A survey. Artificial intelligence in medicine 2019;.

[306] Simonyan K, Zisserman A. Very deep convolutional networks for large-scale image recognition. arXiv preprint arXiv:14091556 2014;.